\documentclass[aps,prc,twocolumn,showpacs,floatfix,nofootinbib,preprintnumbers,superscriptaddress,amsmath,amssymb,longbibliography,amsproc]{revtex4-1}
\usepackage{epsfig,dsfont,amssymb,amsmath,amsthm,amsfonts,amsbsy,mathrsfs}
\usepackage{graphicx}
\usepackage{amsmath}
\usepackage{amssymb}%
\usepackage{dcolumn}
\usepackage{multirow}
\usepackage[colorlinks=true,linkcolor=blue,citecolor=blue,urlcolor=blue]{hyperref}
\usepackage{bbold}
\usepackage{bibentry}
\usepackage[normalem]{ulem}

\graphicspath{{.}{./Figures/}}

\usepackage{stackengine}

\usepackage[disable]{todonotes}

\newcommand{\be}{\begin{equation}}
\newcommand{\ee}{\end{equation}}
\newcommand{\ba}{\begin{eqnarray}}
\newcommand{\ea}{\end{eqnarray}}

\stackMath
\newcommand\tsup[2][2]{%
 \def\useanchorwidth{T}%
  \ifnum#1>1%
    \stackon[-.5pt]{\tsup[\numexpr#1-1\relax]{#2}}{\scriptscriptstyle\sim}%
  \else%
    \stackon[.5pt]{#2}{\scriptscriptstyle\sim}%
  \fi%
}

\DeclareMathAlphabet{\mathpzc}{OT1}{pzc}{m}{it}

\begin{document}

\title{Magnetic dipole transition in $^{48}$Ca}

\thanks{This manuscript has been authored in part by UT-Battelle, LLC, under contract DE-AC05-00OR22725 with the US Department of Energy (DOE). The US government retains and the publisher, by accepting the article for publication, acknowledges that the US government retains a nonexclusive, paid-up, irrevocable, worldwide license to publish or reproduce the published form of this manuscript, or allow others to do so, for US government purposes. DOE will provide public access to these results of federally sponsored research in accordance with the DOE Public Access Plan (http://energy.gov/downloads/doe-public-access-plan).}

\author{B.~Acharya}
\affiliation{Physics Division, Oak Ridge National Laboratory, Oak Ridge, TN 37831, USA} 

\author{B. S.~Hu} 
\affiliation{National Center for Computational Sciences, Oak Ridge National Laboratory, Oak Ridge, TN 37831, USA}
\affiliation{Physics Division, Oak Ridge National Laboratory, Oak Ridge, TN 37831, USA}

\author{S.~Bacca} 
\affiliation{Institut f\"ur Kernphysik and PRISMA+ Cluster of Excellence, Johannes Gutenberg-Universität Mainz, 55128 Mainz, Germany}
\affiliation{Helmholtz-Institut Mainz, Johannes Gutenberg-Universit\"at Mainz, D-55099 Mainz, Germany}

\author{G.~Hagen} 
\affiliation{Physics Division, Oak Ridge National Laboratory, Oak Ridge, TN 37831, USA} 
\affiliation{Department of Physics and Astronomy, University of Tennessee, Knoxville, TN 37996, USA}

\author{P.~Navr\'atil}
\affiliation{TRIUMF, Vancouver, British Columbia V6T 2A3, Canada}

\author{T.~Papenbrock} 
\affiliation{Department of Physics and Astronomy, University of Tennessee, Knoxville, TN 37996, USA} 
\affiliation{Physics Division, Oak Ridge National Laboratory, Oak Ridge, TN 37831, USA}

\begin{abstract} 
The magnetic dipole transition strength $B(M1)$ of  $^{48}$Ca is dominated by a single resonant state at an excitation energy of $10.23$~MeV. Experiments disagree about $B(M1)$ and this impacts our understanding of spin flips in nuclei. We performed {\it ab initio} computations based on chiral effective field theory and found that $B(M1:0^+\rightarrow1^+)$ lies in the range from $7.0$ to $10.2~\mu_N^2$. This is consistent with a $(\gamma,n)$ experiment but larger than results from $(e,e^\prime)$ and $(p,p')$ scattering. Two-body currents yield no quenching of the $B(M1)$ strength and continuum effects reduce it by about 10\%. For a validation of our approach, we computed magnetic moments in $^{47,49}$Ca and performed benchmark calculations in light nuclei. 

\end{abstract}

\maketitle

{\it Introduction.---} Magnetic dipole ($M1$) transitions play an important role in nuclear physics because they probe the spin-isospin structure of nuclei~\cite{heyde2010,neumanncosel2019}. 
They are also important in nuclear astrophysics because of their relation to neutrino-nucleus cross sections~\cite{balasi2015,tornow2022} and other weak processes~\cite{rapaport1994}.  The transition strength $B(M1)$ of $^{48}$Ca is particularly interesting because it constrains electron capture~\cite{luttge1996} in nearby elements that are abundant in the collapsing cores of supernovae~\cite{brown1982}; it  also affects radiative neutron capture processes that set the initial conditions for the core collapse~\cite{loens2012}.

The $B(M1)$ strength distribution in $^{48}$Ca is largely concentrated in the excitation of a single resonant state at excitation energy $E_x=10.23$~MeV~\cite{steffen1980,mathy2017}. This excitation was discovered in inelastic electron scattering experiments by \textcite{steffen1980} and  $B(M1)=4.0\pm 0.3~\mu_N^2$ and $B(M1)=3.9\pm 0.3~\mu_N^2$~\cite{steffen1983} were found. Proton inelastic scattering experiments confirmed these values~\cite{berg1982,rehm1982,fujita1982}, which are much smaller than the estimate based on the extreme single-particle model limit of $12~\mu_N^2$ from a neutron particle-hole excitation $\nu f_{5/2} (\nu f_{7/2})^{-1}$. In contrast, the more recent $^{48}\mathrm{Ca}(\gamma,n)$ experiment by \textcite{tompkins2011} reported the value $B(M1)=6.8\pm 0.5~\mu_N^2$. The subsequent analysis of proton scattering reactions~\cite{birkhan2016} extracted $B(M1)=3.45(85)$ to $4.1(1.0)~\mu_N^2$, which is consistent with the older values. 
The incompatibility of the experimental values from $(e,e^\prime)$ and $(p,p^\prime)$ scattering with the $(\gamma,n)$ reaction is puzzling. 

\begin{figure}[htb]
    \includegraphics[width=0.49\textwidth]{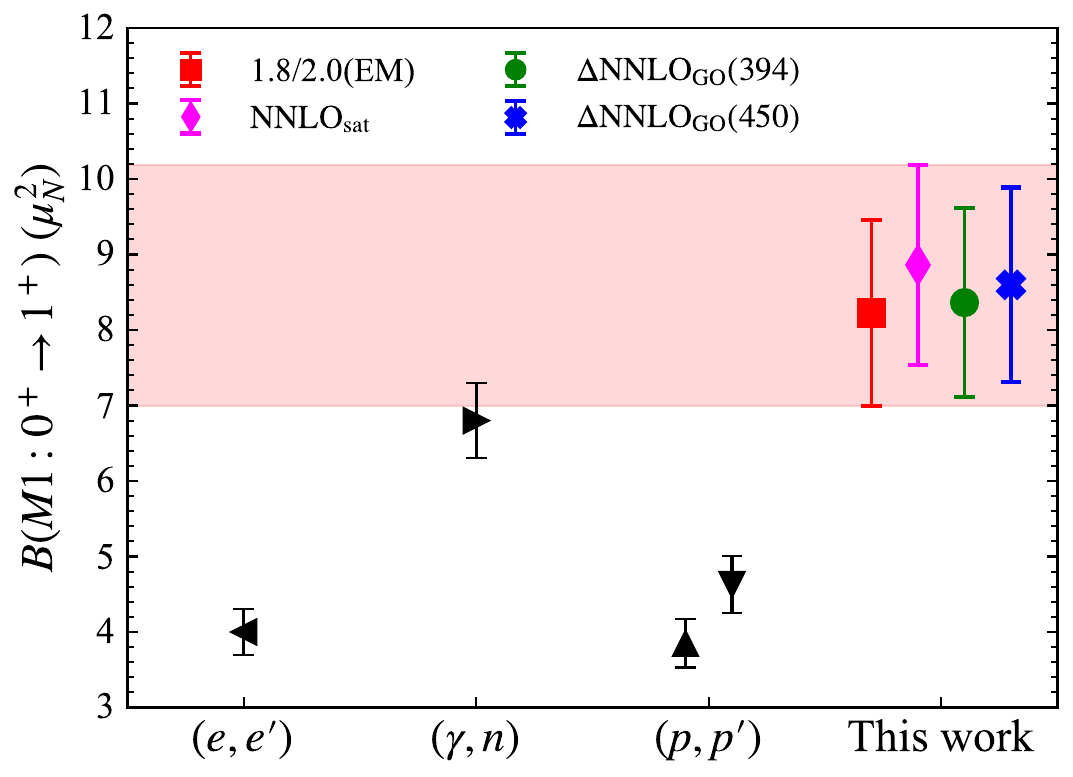}
    \caption{The $B(M1)$ strength carried by the $1^+$ state at $10.23$~MeV in $^{48}$Ca. Data from the $(e,e^\prime)$~\cite{steffen1980}, $(\gamma,n)$~\cite{tompkins2011}, and $(p,p')$~\cite{birkhan2016} experiments are compared to the calculations of this work.} 
    \label{fig:ca48_bm1}
\end{figure}

Theory has struggled to explain a small $M1$ strength in $^{48}$Ca~\cite{mcgrory1981,kohno1982,takayanagi1987,brand1990,brown1998,holt2014}. The early~\cite{mcgrory1981} and later~\cite{brown1998} shell-model calculations showed that correlations reduce the single-particle estimate to about $B(M1)=7.4~\mu_N^2$, quite above the results from $(e,e^\prime)$ and $(p,p^\prime)$ scatterings. More recent shell-model calculations based on Hamiltonians from chiral effective field theory including two- and three-nucleon forces reported  $B(M1)\approx 5.3~\mu_N^2$~\cite{holt2014}. 
Early on, meson exchange currents were thought to significantly reduce the strength~\cite{suzuki1981,harting1981,rehm1982,steffen1983}, though there were doubts that the resulting reduction could be large enough to explain the  data~\cite{bertsch1981,kohno1982}. 
Modern shell-model calculations often employ the quenching factor $q=0.75$ to include the effect of meson-exchange currents and thereby reduce the computed $B(M1)$ by a factor of $q^2\approx 0.56$ to about $4.7$ to $5.5~\mu_N^2$~\cite{holt2014,wilhelmy2018}. 
This quenching factor was found by comparing $M1$ strength distributions from $(e,e^\prime)$ scattering with shell-model calculations~\cite{neumanncosel1998}. It is close to the  empirical quenching of the Gamow-Teller strength~\cite{brown1985,martinez1996} in the nuclear shell model. However, while the isovector part $\sigma \tau_0$ of the $M1$ operator and the Gamow-Teller operators $\sigma \tau_\pm $ are isospin components of an isovector operator, the two-body currents that quench Gamow-Teller decays~\cite{menendez2011,gysbers2019} are different from those for the $M1$ transitions~\cite{pastore2009,piarulli2013,pastore2013,gnech2022,pal2023}. 
We note finally that the excited $1^+$ state at 10.23~MeV in $^{48}$Ca is a resonance that decays by neutron emission. As the  amplitude of this (not square integrable) state enters the $M1$ transition matrix element, one wonders how continuum effects impact the $B(M1)$. In this Letter, we present a calculation that takes into account all of these effects: wave function correlations, two-body currents, and continuum features. Our final result is compared to the experiments in Fig.~\ref{fig:ca48_bm1}.

{\it Methods.---}We employed nucleon-nucleon and three-nucleon forces as well as two-body currents from chiral effective field theory~\cite{epelbaum2009}. Earlier studies along these lines were carried out for the $M1$ moments~\cite{pastore2009,piarulli2013,gnech2022,martin2023,pal2023,seutin2023} and also for the transition strengths~\cite{pastore2013,navratil2007b} of light nuclei. We employed the coupled-cluster method to compute the ground and the excited $1^+$ states in $^{48}$Ca, and the $M1$ transition matrix element.

We used the intrinsic Hamiltonian
\begin{equation}
    H = T-T_{\rm CM} +V_{NN} +V_{NNN} \ .
\end{equation}
Here, $T$ and $T_{\rm CM}$ denote the kinetic energy of $A$ nucleons and of the center of mass, respectively. The nucleon-nucleon potential $V_{NN}$ and three-nucleon potential $V_{NNN}$ are from chiral effective field theory. Our calculations use the potentials 1.8/2.0(EM) of Ref.~\cite{hebeler2011}, NNLO$_{\rm sat}$ of Ref.~\cite{ekstrom2015a}, and the $\Delta$NNLO$_{\rm GO}(394)$ and $\Delta$NNLO$_{\rm GO}(450)$ of Ref.~\cite{jiang2020}. The potential 
1.8/2.0(EM) is based on a similarity renormalization group transformation~\cite{bogner2007} of the nucleon-nucleon potential by \textcite{entem2003} and includes the leading three-nucleon forces from chiral effective field theory. The NNLO$_{\rm sat}$ potential consists of nucleon-nucleon and three-nucleon potentials at next-to-next-to-leading order in chiral effective field theory. The $\Delta$NNLO$_{\rm GO}$ potentials consist of nucleon-nucleon and three-nucleon potentials at next-to-next-to-leading order in chiral effective field theory with $\Delta$ isobars included. 
They have cutoffs of 394 and 450~MeV. The difference of their results shown in Fig.~\ref{fig:ca48_bm1} yields a 5\% estimate for the cutoff dependence.

For the computation of $M1$ matrix elements,  we employed current operators derived in chiral effective field theory. The leading order one-body operator is given by~\cite{Park:1995pn,pastore2009,Schiavilla:2018udt}
\begin{equation}
    \label{eq:mu1b}
    \boldsymbol{\mu}_\mathrm{LO} = \frac{1}{2}\sum_{i=1}^A \mu_n \boldsymbol{\sigma}_i [1+\tau_i^{(z)}] + 
    (\ell_i + \mu_p \boldsymbol{\sigma}_i) [1-\tau_i^{(z)}]\ .
\end{equation}
Here $\ell_i$ and $\boldsymbol{\sigma}_i$ are the orbital and Pauli spin operators acting on the nucleon $i$, and 
the isospin operators $[1\pm\tau_i^{(z)}]/2$ project the nucleon to neutron/proton states with magnetic moments $\mu_{n,p}$. At next-to-leading-order, $\boldsymbol{\mu}_\mathrm{NLO} = \boldsymbol{\mu}_\mathrm{intr}+\boldsymbol{\mu}_\mathrm{Sachs}$~\cite{pastore2009,seutin2023,Ericson:1988gk} with
\begin{align}
    \label{eq:intr}
    \boldsymbol{\mu}_\mathrm{intr} = i e \frac{g_A^2 m_\pi}{64\pi f_\pi^2} \sum_{i,j>i}^A & \Big[\Big(\frac{1}{ r^2}+\frac{1}{m_\pi r^3}\Big)
    (\boldsymbol{\sigma}_i \times \boldsymbol{\sigma}_j)\cdot\mathbf{r}\,\mathbf{r}  \nonumber\\
    & \, -(\boldsymbol{\sigma}_i \times \boldsymbol{\sigma}_j)\Big]e^{-m_\pi r}\,\tau_{ij}^{(\times)}\,,
\end{align}
and 
\begin{equation}
    \label{eq:sachs}
    \boldsymbol{\mu}_\mathrm{Sachs} = \frac{i}{4} \, e \sum_{i,j>i}^A V_{1\pi}(r) \,\mathbf{R}\times\mathbf{r}\,\tau_{ij}^{(\times)}\,.
\end{equation}
Here $\mathbf{r} = \mathbf{r}_i-\mathbf{r}_j$, $\mathbf{R} = (\mathbf{r}_i+\mathbf{r}_j)/2$, and $V_{1\pi}(r)$ is the leading-order one-pion-exchange nucleon-nucleon potential. As in the potentials used in this work, we assigned the shifted value $g_A=1.29$ to the axial coupling constant in $V_{1\pi}(r)$ and in Eq.~\eqref{eq:intr} to satisfy the Goldberger-Treiman relation. Consistent with our convention for the nucleon isospin projections as shown in Eq.~\eqref{eq:mu1b}, we have $\tau_{ij}^{(\times)}\equiv-2i(\boldsymbol{\tau}_i \times \boldsymbol{\tau}_j)_z$ for the isospin operator. We followed Particle Data Group~\cite{Workman:2022ynf} recommendations for the values of $\mu_{n,p}$, the pion decay constant $f_\pi$ and the charged pion mass $m_\pi$. 
The magnetic moment operators in Eqs.~\eqref{eq:intr} and ~\eqref{eq:sachs} exhibit unproblematic $1/r$ and $1/r^2$ singularities, respectively, and do not require a cutoff for regularization.

{\it Computation of the transition matrix element.---}The coupled-cluster computations~\cite{coester1958,coester1960,kuemmel1978,bishop1991,bartlett2007,hagen2014} of the groundstate and the excited $1^+$ state were performed in the CCSDT-3~\cite{noga1987} and EOM-CCSDT-3~\cite{watts1996} approximations, respectively. The latter includes singles (one-particle--one-hole), doubles (two-particle--two-hole), and the most relevant triples (three-particle--three-hole) excitations. We briefly summarize the well-documented steps~\cite{bartlett2007,hagen2014}. The calculations used a spherical harmonic oscillator basis consisting of $N_{\rm max}+1$ shells with a spacing of $\hbar\omega$. Due to the computational cost of EOM-CCSDT-3, our calculations were limited to $N_{\rm max} = 12$. We gauged the model-space convergence by varying $\hbar\omega$ in the range $14$--$20$~MeV, and by performing EOM-CCSD calculations for $N_{\rm max} = 12, 14$. Based on the observed convergence, we selected $N_{\rm max} = 12$ and  $\hbar \omega= 16$~MeV, and included an estimated residual model-space uncertainty in the total estimated uncertainty. Starting from a spherical Hartee-Fock reference state, we solved the coupled-cluster equations and computed the similarity transformed Hamiltonian. We solved for the left groundstate and excited $1^+$ state via equations-of-motion techniques, and the transition matrix element was computed as usual in a bi-variational framework~\cite{bartlett2007}. We normal-ordered the two-body currents with respect to the Hartree-Fock reference state and included the normal-ordered one-body terms in the computation of the transition matrix elements. 
We verified that the residual normal-ordered two-body part gives only a small contribution to the magnetic moments of odd calcium isotopes; presumably, their effect on the transition strength is also small~\cite{gysbers2019}.
To account for the fact that the $1^+$ excited state is a resonant state, we computed the transition to the $0^+$ ground-state in the EOM-CCSD approximation using a Berggren basis~\cite{berggren1968,berggren1971} for the $\nu f_{5/2}$ partial wave (see Ref.~\cite{hagen2016} for details). Including continuum effects in this partial wave is sufficient as our computations revealed that, for the 1.8/2.0(EM) interaction, one-particle--one-hole excitations from the $\nu f_{7/2}$ to the $\nu f_{5/2}$ partial wave account for more than 90\% of the norm of the $1^+$ state.

For all interactions considered in this work, the coupled-cluster calculations yielded a single $1^+$ state near the neutron-separation threshold. As reported in Table~\ref{tab1}, the 1.8/2.0(EM) interaction gives a resonant state at about 0.6~MeV above the neutron-emission threshold while the other interactions give a weakly bound $1^+$ state. We identify this state with the experimentally known $1^+$ resonance at 0.28~MeV above the neutron-emission threshold. As the one-particle--one-hole excitations account for more than 90\% of the norm of this state, we are confident that our EOM-CCSDT-3 calculations were well converged with respect to neglected higher-order excitations.  We finally note that the computed resonant width for the potential 1.8/2.0(EM) agrees with results from a toy model where we employed a Woods-Saxon potential (with standard parameters) such that the $\nu f_{7/2}$ groundstate is bound by the neutron separation energy and the $\nu f_{5/2}$ state is a resonance at the experimental energy.

\begin{table}[!htb]
 \centering

 \begin{tabular}{|l r r c c |}
 \hline
\hline
 &~~~~$S_n$ &~~~~$~\Delta E~$ & $~~~\Gamma~$ & $1p$--$1h$ \\

 &~~~~(MeV)   &~~~~(MeV)~&~~~~(keV) & ~ \\
\hline

     $\Delta$NNLO$_\mathrm{GO}$(394)     & 9.74    & $-0.44 $ & 0 & ~91\% \\
     $\Delta$NNLO$_\mathrm{GO}$(450)     &  9.38   & $-1.26 $ & 0 & ~91\% \\
     NNLO$_\mathrm{sat}$                 &  9.34   & $-0.23 $ & 0 & ~91\% \\
     1.8/2.0(EM)                         &  10.00  & $~0.55 $ & 4 & ~92\% \\
     Experiment                          &  9.95   & $~0.28$ & $\leq17$ & \\
 \hline
 \hline
\end{tabular}
\caption{The neutron-separation energy $S_n$, the energy $\Delta E$ of the $1^+$ state with respect to the neutron-emission threshold, the resonance width $\Gamma$ of $^{48}$Ca, and the $1p$--$1h$ component of the norm of the $1^+$ state for the potentials as indicated and compared to experiment. The data for $\Gamma$ is taken from Ref.~\cite{tamii2007}.}
\label{tab1}
\end{table}

\begin{figure}[htb]
    \includegraphics[width=0.49\textwidth]{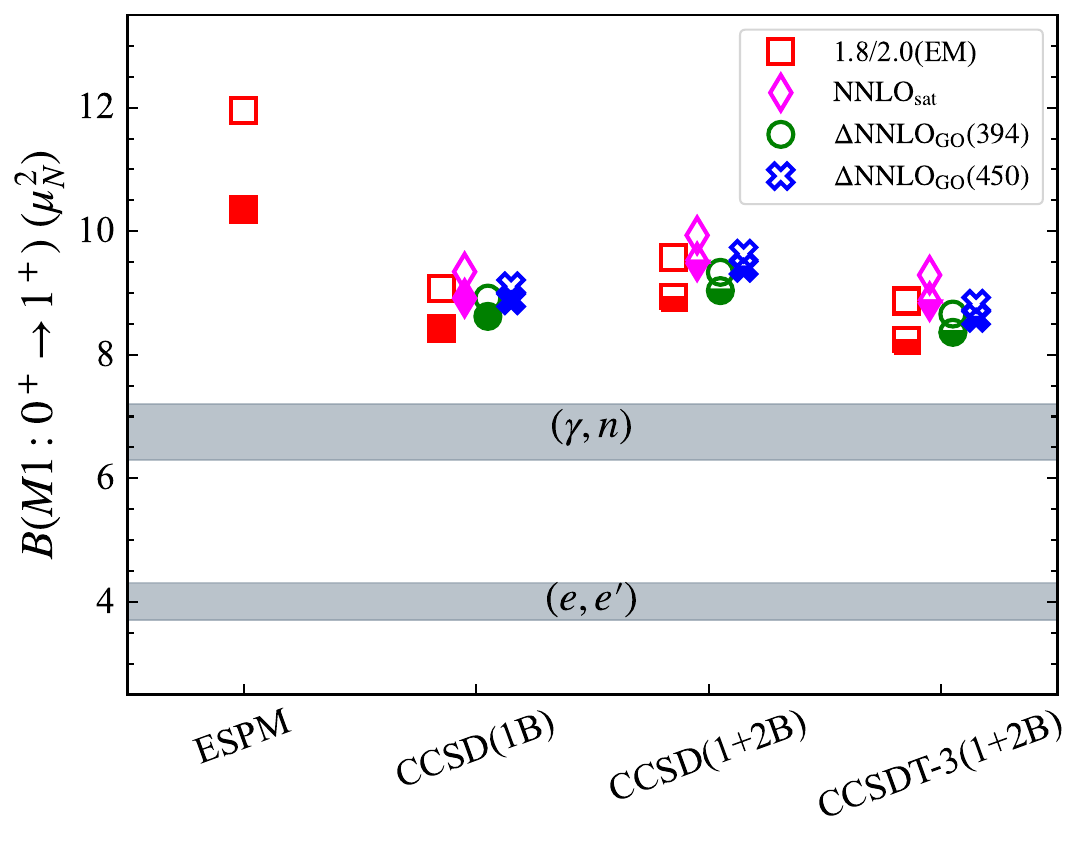}
    \caption{The $B(M1)$ computed in a harmonic oscillator basis (empty symbols) and a Gamow basis (filled symbols) for different potentials and methods. The extreme single-particle model (ESPM) shows the impact of the continuum. Half-filled symbols are from the harmonic oscillator basis where continuum effects, estimated from the CCSD calculations with one-body currents (1B), were added to the CCSD and CCSDT-3 calculations that use the one-body and the normal-ordered one-body parts of the two-body currents (1+2B). The bands show results from the $(e,e^\prime)$~\cite{steffen1980} and $(\gamma,n)$~\cite{tompkins2011} experiments.}  
    \label{fig:bm1_detailed}
\end{figure}

{\it Composition of the $B(M1)$ strength.---} 
Figure~\ref{fig:bm1_detailed} shows how correlations, two-body currents, and the continuum impact the $B(M1)$ strength for the $1^+$ state in $^{48}$Ca. We start with the extreme single-particle model. The maximum value of $B(M1)\approx 12~\mu_N^2$ from the spin-orbit partners in the harmonic oscillator is significantly reduced to about 10.4 when a Gamow-Hartree-Fock basis~\cite{hagen2006b} is used for the $\nu f_{5/2}$ partial wave. We checked that this value is consistent with the toy model. This shows how the particle continuum reduces the $M1$ strength. The CCSD approximation includes up to two-particle--two-hole excitations and further reduces the $M1$ strength. Continuum effects, computed in the CCSD approximation with the  $\nu f_{5/2}$ partial wave in a Gamow basis, further reduce the $M1$ strength. The inclusion of two-body currents increases the $M1$ strength. This finding is consistent with the results reported in Refs.~\cite{Park:1995pn,piarulli2013,phillips2016} for $A\leq 3$ nuclei, in Ref.~\cite{friman-gayer2021} for $^6$Li, in Ref.~\cite{marcucci2008} for $A=6,7$ nuclei, and in Ref.~\cite{pastore2013} for $6<A<9$ nuclei. Clearly, the two-body currents impact the $B(M1)$  quite differently than Gamow-Teller decays (where they produce quenching~\cite{gysbers2019}). This
is contrary to the expectation of Refs.~\cite{heyde2010,birkhan2016}. We also note that the $\Delta$-excitation currents, to which strong quenching of $B(M1)$ of $^{48}$Ca has traditionally been attributed~\cite{harting1981,steffen1983,kohno1982}, are of higher order in chiral effective field theory than the leading one-pion-exchange currents~\cite{Schiavilla:2018udt}. The calculations of Refs.~\cite{fabian1976,marcucci2008,gnech2022} found that the effects of the $\Delta$-excitation currents are indeed small compared to the leading one-pion-exchange contributions at low energies. Returning to Fig.~\ref{fig:bm1_detailed}, the inclusion of triples, i.e., three-particle--three-hole excitations, reduces the $M1$ strength. Our coupled-cluster calculations where the $\nu f_{5/2}$ is from a Gamow basis are limited to the CCSD approximation and one-body currents. The half-filled symbols are obtained by simply adding that reduction to the calculations with two-body currents and triples correlations.    

\begin{figure*}[htbp]
    \includegraphics[width=0.9\textwidth]{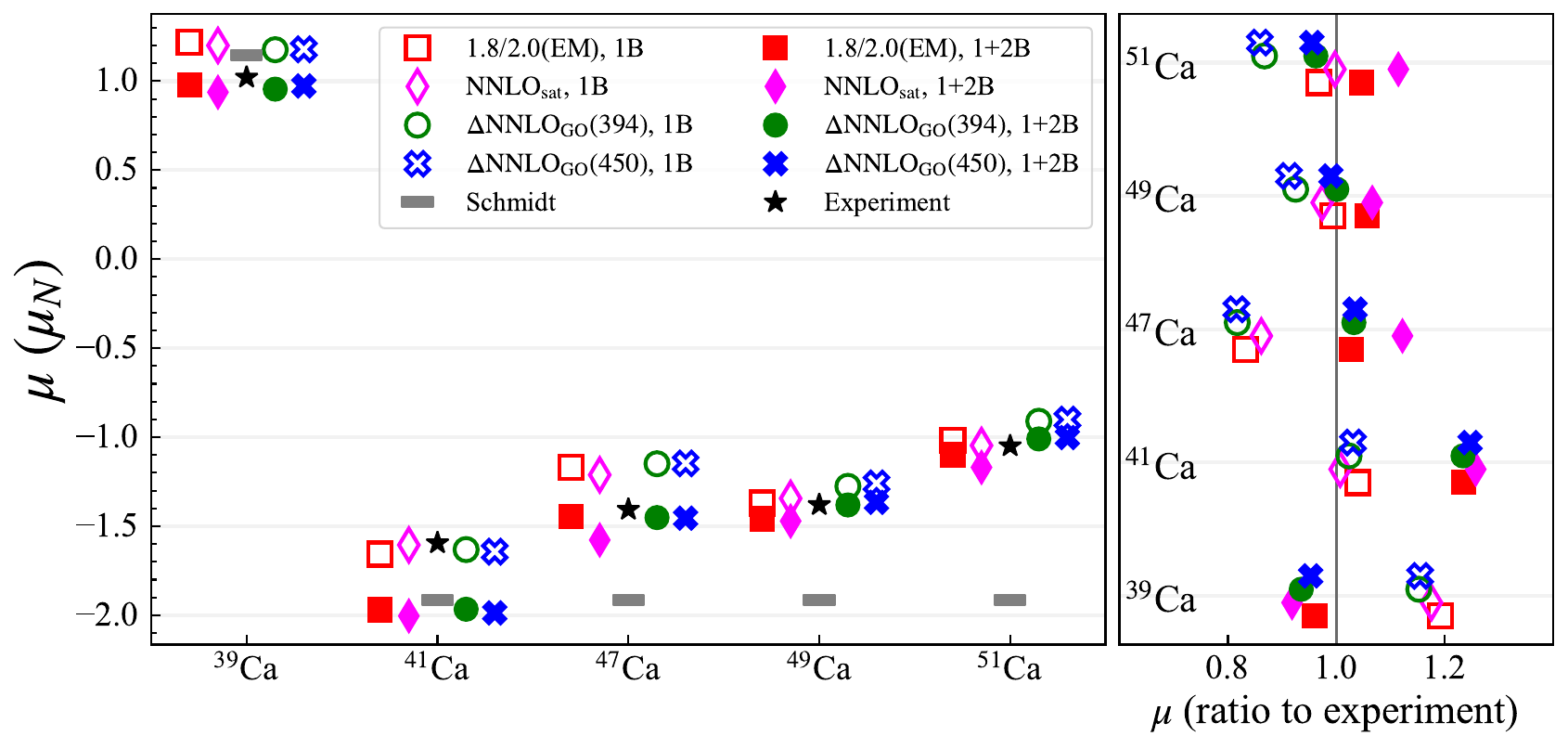}
    \caption{Magnetic moments of odd-mass neighbors of the doubly-magic nuclei $^{40,48}$Ca (left) from interactions as indicated. The hollow symbols result from one-body currents only while the full symbols include two-body currents. Experimental values are taken from Refs.~\cite{minamisono1976,brun1962,ruiz2015} and the Schmidt values are based on the extreme single-particle model. The right panel shows the ratio of computed to measured magnetic moments. }
    \label{fig:ca_moments}
\end{figure*}

Let us discuss the uncertainties of the $B(M1)$ computation. The $B(M1)$ predictions for the four different chiral interactions agree within 7\%. We estimate that model-space uncertainties (from varying the oscillator frequency $\hbar\omega$ and increasing $N_{\rm max}$) amount to about 3\% of the $B(M1)$ value. The Gamow basis virtually eliminates infrared corrections, and this makes the model-space uncertainty small. Going from the doubles (two-particle--two-hole) to the triples (three-particle--three-hole) approximation decreases the $B(M1)$ results by about 10\%. The inclusion of two-body currents increases the $M1$ strength by about 5\%. Thus, we give a conservative uncertainty estimate of 15\% on the computed $B(M1)$. 
Calculations in $^{14,22}$O using different many-body methods (for the same interaction~\cite{soma2020}) present us with another way to validate our calculations and uncertainty estimates. These nuclei are similar to $^{48}$Ca because the $1^+$ state is a one-particle--one-hole excitation between neutron spin-orbit partners.  In $^{22}$O, the $B(M1)$ strength from coupled-cluster theory is about 10\% lower than the no-core shell model~\cite{barrett2013,navratil2016} with importance truncation~\cite{roth2008a}. In $^{14}$O, both calculations are not fully converged with respect to increasing the model space, and one would estimate that a further 15\% reduction in strength is reasonable. Based on these benchmarks (see Supplemental Material~\cite{suppl_M1}) we would again assign an uncertainty of about 15\% to the computed $B(M1)$ in $^{48}$Ca. This uncertainty is shown in Fig.~\ref{fig:ca48_bm1} for each of the potentials and the overall uncertainty encompasses all. This yields a $B(M1)$ value in the range from $7.0$ to $10.2~\mu_N^2$.        

As a check, we also computed the $B(M1)$ using the form factor for the transition between the $0^+$ groundstate and the $1^+$ excited state. 
This form factor is proportional to the reduced matrix element of $T_1^\mathrm{mag}(q)$, the transverse magnetic multipole~\cite{acharya2020} of the $q$-dependent electromagnetic current operator~\cite{pastore2009} for multipolarity~1. It is related to $B(M1)$ via 
\begin{equation}
   \lim_{q \rightarrow E_x(1^+)}  \frac{4\pi}{Z^2q^2}  \vert \langle 1^+ \vert \vert T_1^\mathrm{mag}(q) \vert \vert 0^+ \rangle \vert^2 = \frac{8\pi}{9Z^2}  B(M1)\,.
   \label{eq:ff}
\end{equation}
This  calculation of the $B(M1)$ agreed with the direct computation (see Supplemental Material~\cite{suppl_M1}).

{\it $M1$ moments of Ca isotopes.---}
For further validation, we computed magnetic moments in the odd calcium isotopes $^{39,41,47,49,51}$Ca in the CCSD approximation using axially symmetric (and nearly spherical) reference states (see Ref.~\cite{novario2020} for details. 
Here we included both normal-ordered one- and two-body parts of the two-body currents. In Fig.~\ref{fig:ca_moments}, the hollow symbols show the results when only one-body currents are used, while the filled symbols also include the two-body currents. We see that the inclusion of two-body currents typically moves the results closer to data -- with $^{41}$Ca being the exception. This is unexpected. Comparison with the results from \textcite{Miyagi2023} shows that their single-shell results are close to ours for all nuclei while their multi-shell results are closer to data for $^{41}$Ca (albeit further away for lighter calcium isotopes). With a view on Ref.~\cite{Miyagi2023}, we also note that angular-momentum projection techniques~\cite{hagen2022} and computations of heavier nuclei are under development within coupled-cluster theory.

To validate our calculations of magnetic moments, we also computed magnetic moments (using one-body currents) in $^{15,17}$O and compared them with the importance-truncated no-core shell model~\cite{roth2008a}, using the interaction of Ref.~\cite{soma2020}. The calculations agree to within a fraction of 1\% (see Supplemental Material~\cite{suppl_M1}).

{\it Discussion.---} 
Our theoretical results are about a factor of two larger than what is extracted from $(e,e^\prime)$ and $(p,p^\prime)$ scattering but agree within uncertainties with the $(\gamma,n)$ experiment. It is not easy to see what could introduce a significant further reduction of the computed strength. Regarding wave function correlations, the energy and structure of the $1^+$ excited state seem sufficiently simple and well converged after three-particle--three-hole excitations are included, and their effect is small compared to the dominant one-particle--one-hole excitations. Continuum effects are important as well and they have been included in this work.  While there are higher-order two-body currents in the effective field theory, those are expected to be smaller than the leading contributions we employed; this has indeed been found in magnetic properties of light nuclei~\cite{piarulli2013,gnech2022}.

{\it Summary.---}
We computed the magnetic dipole strength of the $1^+$ excited state in $^{48}$Ca at 10.23~MeV using Hamiltonians and currents from chiral effective field theory. Our calculations included continuum effects, three-nucleon forces, and two-body currents. The latter differ from those that enter Gamow-Teller transitions and yield only a smaller contribution to the $M1$ strength. Our results agree within uncertainty estimates with those from a $(\gamma,n)$ measurement but are significantly above those from $(e,e^\prime)$ and $(p,p^\prime)$ scattering. A clarification of the experimental situation could potentially inform and impact estimates for neutrino-nucleus cross sections and {\it ab initio} computations based on chiral effective field theory.
\newline{}  

\begin{acknowledgments}
We thank Takayuki Miyagi for sharing unpublished results with us. We also thank Mitch Allmond, George Bertsch, Dick Furnstahl, Tim Gray, Augusto Macchiavelli, Takayuki Miyagi, and Achim Schwenk for discussions, and Peter von Neumann-Cosel and Achim Richter for communications. This work was supported by the U.S. Department of Energy, Office of Science, Office of Nuclear Physics, under Award Nos.~DE-FG02-96ER40963 and the SciDAC-5 NUCLEI collaboration, and by the Office of High Energy Physics, U.S. Department of Energy under Contract No. DE-AC02-07CH11359 through the Neutrino Theory Network Fellowship awarded to B.~Acharya. This work was supported  by the Deutsche Forschungsgemeinschaft (DFG) through the Cluster of Excellence ``Precision Physics, Fundamental Interactions, and Structure of Matter" (PRISMA$^+$ EXC 2118/1) funded by the DFG within the German Excellence Strategy (Project ID 39083149). This work was supported by the NSERC Grant No. SAPIN-2022-00019. TRIUMF receives federal funding via a contribution agreement with the National Research Council of Canada. Computer time was provided by the Innovative and Novel Computational Impact on Theory and Experiment (INCITE) program. This research used resources from the Oak Ridge Leadership Computing Facility located at ORNL, which is supported by the Office of Science of the Department of Energy under Contract No. DE-AC05-00OR22725, and from the Digital Research Alliance of Canada.
\end{acknowledgments}

\bibliography{main}
\clearpage
\onecolumngrid
\setcounter{page}{1}
\setcounter{equation}{0}
\setcounter{figure}{0}

\section*{Supplemental Material: The magnetic dipole transition in $\mathbf{^{48}Ca}$}

\subsection{The transition form factor}

\begin{figure}[h]
    \includegraphics[width=0.8\textwidth]{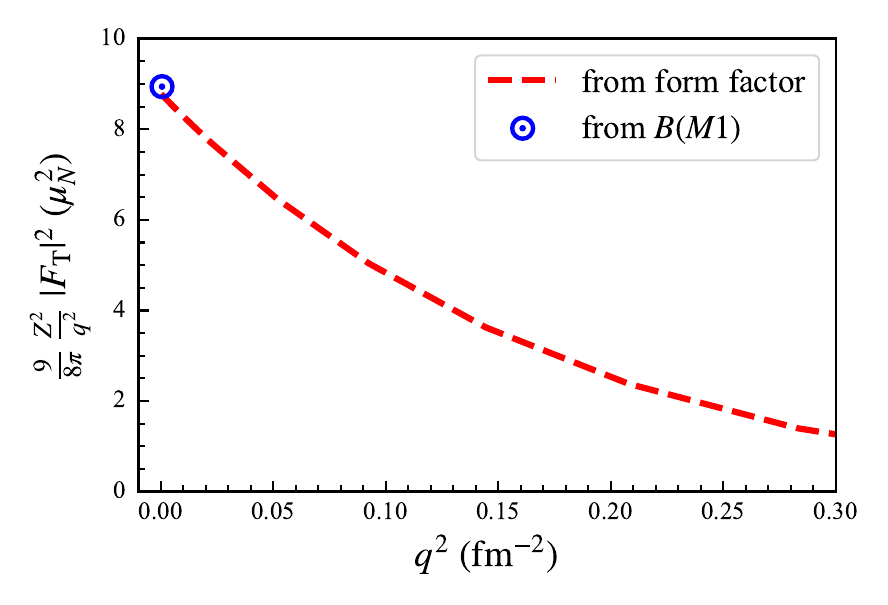}
    \caption{$M1$ transition form factor (red dashed line) where its zero four-momentum transfer limit agrees with the direct calculation of the $B(M1)$ (blue bull's eye).}
    \label{fig:bm1_from_ff}
\end{figure}

Figure~\ref{fig:bm1_from_ff} compares the transition form factor $F_\mathrm{T}$ at $q=E_x(1^+)$ obtained by evaluating $B(M1)$, as the squared reduced matrix element of the magnetic dipole operator between the $0^+$ and the $1^+$ states, with the curve obtained by computing $F_\mathrm{T}$ as a function of $q$ (see Eq.~\eqref{eq:ff}). 
Calculations were performed in the CCSD approximation and were based on the $\Delta$NNLO$_\mathrm{GO}(394)$ interaction in a model space with parameters $N_\mathrm{max}=10$ and  $\hbar\omega=16$, using one-body currents. The two calculations agree.

\subsection{Benchmarks with the no-core shell model}

To validate the uncertainty estimates of  our calculations, we computed the $M1$ transition strengths and moments of oxygen isotopes and compared them with the results from the 
no-core shell model. The benchmarks were performed using the one-body $M1$ operator and the chiral interaction of Ref.~\cite{soma2020} for oscillator energy $\hbar\omega=18$~MeV.  The no-core shell model used the $N\hbar\omega$ basis space along with importance truncation~\cite{roth2008a} for the largest $N$ value.

\subsubsection{$B(M1)$ of $^{14,22}\mathrm{O}$}
\begin{figure}[htb]
    \includegraphics[width=0.9\textwidth]{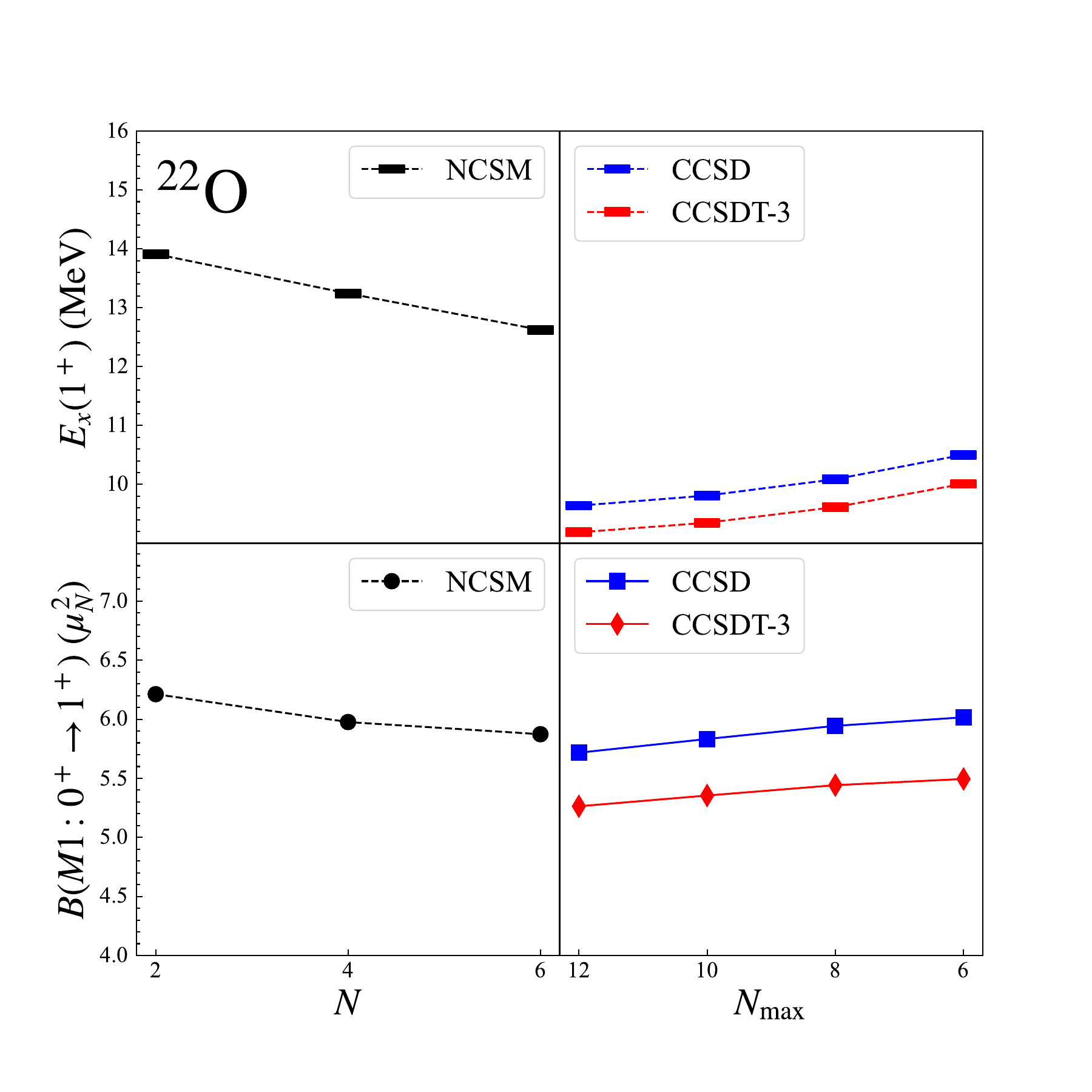}
    \caption{Comparison of the excitation energy $E_x(1^+)$ of the $1^+$ state (top panels) and its $B(M1)$ (bottom panels) in $^{22}$O computed in no-core shell model (left panels) and coupled-cluster theory in the CCSD and CCSDT-3 approximations (right panels) with model space parameters $N$ and $N_\mathrm{max}$, respectively. 
    }
    \label{fig:o22_bm1}
\end{figure}

Figure~\ref{fig:o22_bm1} shows a benchmark for the excitation energy $E_x(1^+)$ of the $1^+$ state and the $B(M1)$ in $^{22}$O. Here the no-core shell model and the CCSDT-3 results differ by about 10\% for the $B(M1)$, while the energy is not yet converged for the no-core shell model.   

\begin{figure}[htb]
    \includegraphics[width=0.9\textwidth]{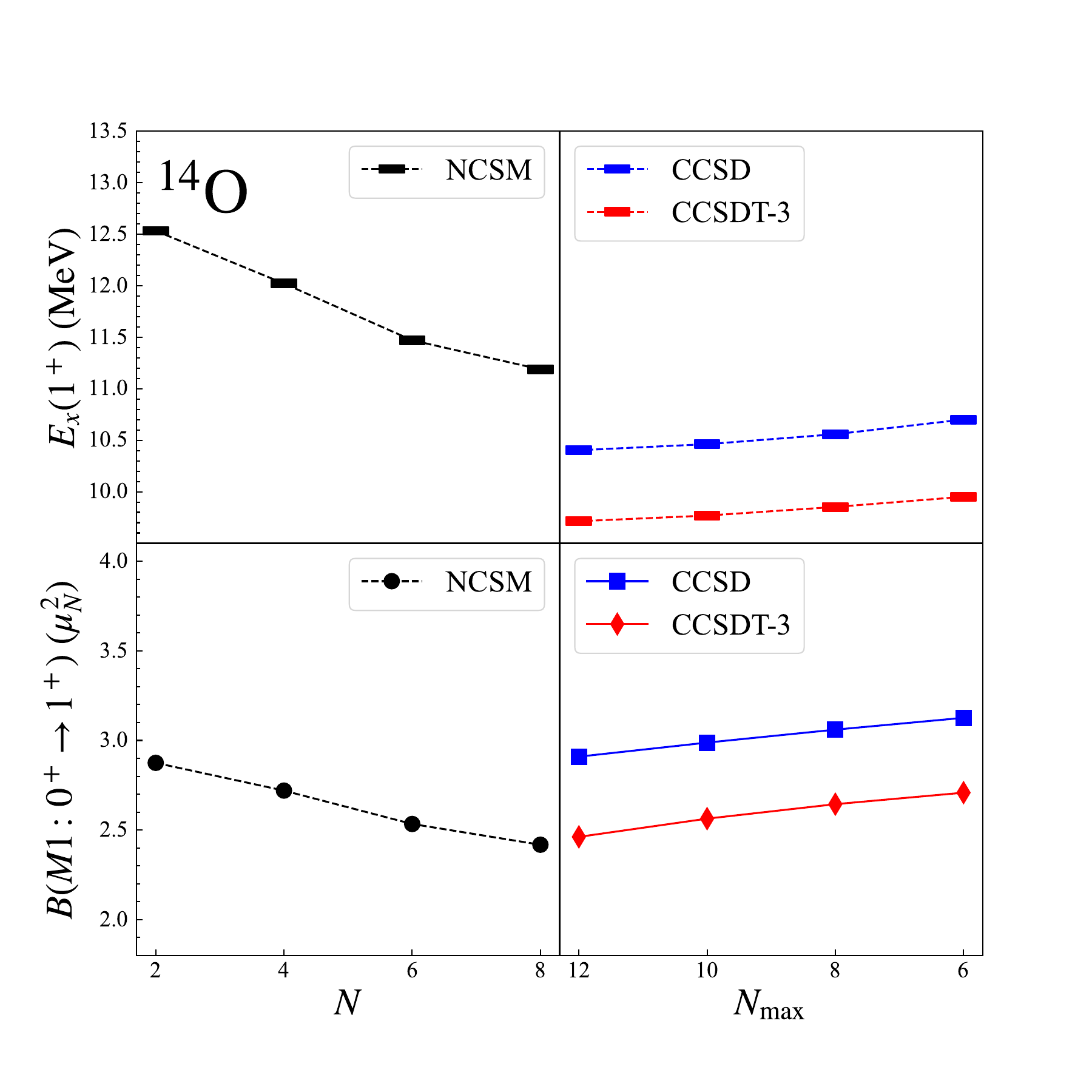}
    \caption{Comparison of the excitation energy $E_x(1^+)$ of the $1^+$ state (top panels) and its $B(M1)$ (bottom panels) in $^{14}$O computed in no-core shell model (left panels) and coupled-cluster theory in the CCSD and CCSDT-3 approximations (right panels) with model space parameters $N$ and $N_\mathrm{max}$, respectively. 
    }
    \label{fig:o14_bm1}
\end{figure}

Figure~\ref{fig:o14_bm1} shows the same results for the nucleus $^{14}$O. Although the no-core shell model and the CCSDT-3 results of $B(M1)$ appear to be approaching similar values, neither is fully converged with respect to the model space. Based on a conservative estimate, we assign an uncertainty of 15\% to our coupled-cluster computations. 

\subsubsection{Magnetic moments of $^{15,17}\mathrm{O}$}

\begin{figure}[htb]
    \includegraphics[width=0.9\textwidth]{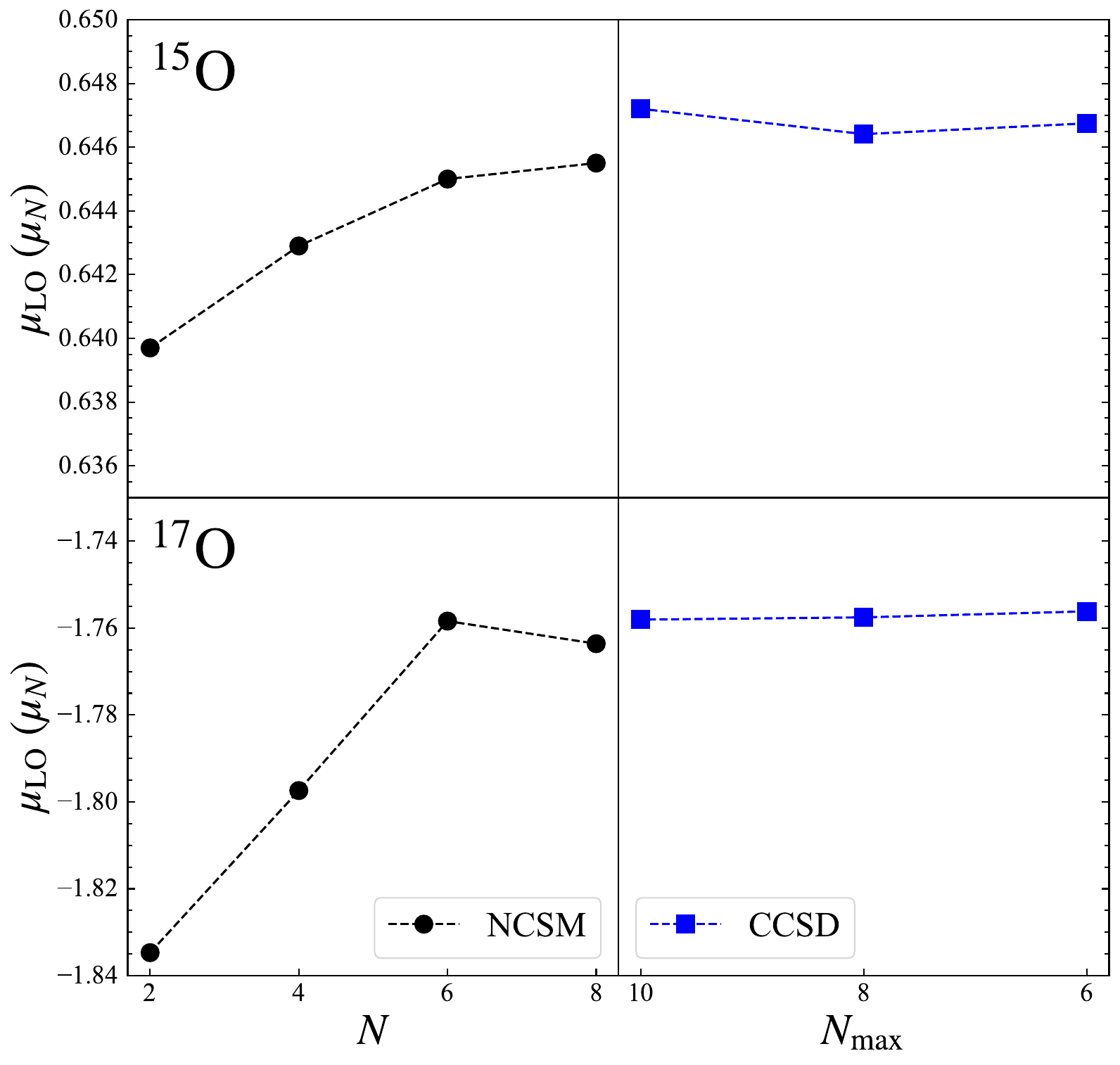}
    \caption{Comparison of the $M1$ moments of $^{15}$O (top panels) and $^{17}$O (bottom panels) computed in no-core shell model (left panels) and coupled-cluster theory in the CCSD approximation (right panels) with model space parameters $N$ and $N_\mathrm{max}$, respectively. 
    }
    \label{fig:o15_17_mom}
\end{figure}

Figure~\ref{fig:o15_17_mom} shows benchmarks for the $M1$ ground-state moments of $^{15,17}\mathrm{O}$. The moments are well converged with respect to the size of the model spaces and differ by less than 1\%. 

\twocolumngrid
\end{document}